\documentclass[pdftex,11pt]{article}
\usepackage{fullpage}
\usepackage{url,amsmath}
\usepackage[dvips]{graphicx}
\usepackage[numbers]{natbib}
\usepackage[
            plainpages=false,
            pdfpagelabels,
            pageanchor,
            pdfstartview=FitH,
            bookmarks=false,
            colorlinks=true,
            citecolor=blue,
            linkcolor=blue,
            ]{hyperref}\pdfoutput=1
            
\hypersetup{
  colorlinks,
  citecolor=blue,
  linkcolor=blue,
  urlcolor=blue}

\begin{document}

\title{Instability of a Flame Front in a Lean H$_2$-N$_2$O Mixture}
\author{Sally P. M. Bane, R\'{e}my M\'{e}vel, Joseph E. Shepherd \\
\\\vspace{6pt} Explosion Dynamics Laboratory, Graduate Aerospace Laboratories \\ California Institute of Technology, Pasadena, CA  91125, USA}
\maketitle

\begin{abstract}
Spark ignition and subsequent flame propagation in a premixed, fuel-lean ($\phi=0.4$) H$_2$-N$_2$O mixture are shown in this fluid dynamics video.  High-speed schlieren photography was used to visualize the flame propagation and the complex instabilities that developed on the flame surface.  The instabilities are a result of the competition between diffusive processes at the flame front and hydrodynamic instability.
\end{abstract}

% main text
\section{Background}

An initially smooth propagating flame front can develop a cellular structure due to two possible types of instability.  The thermodiffusive instability is well known to cause cellular flames in mixtures where the lighter reactant is deficient.  For example, in hydrogen-oxygen mixtures, the hydrogen is lighter in weight than the oxygen and so it diffuses more quickly.  Therefore, if the mixture is fuel-lean, i.e., has less hydrogen than required for a stoichiometric mixture, then the flame will be subject to the thermodiffusive instability.  A flame will be unstable with respect to thermodiffusive effects if the mixture Lewis number is below some critical value \citep{BechtoldMatalon1987}.  The Lewis number is defined as the ratio of the mixture heat diffusivity to the mass diffusivity of the limiting reactant (e.g., hydrogen),
\begin{align}
Le = \frac{\alpha}{D} &= \frac{\kappa/\left(\rho c_p\right)}{D} 
\end{align}
where $\kappa$, $\rho$, and $c_p$ are the thermal conductivity, density, and specific heat capacity of the mixture respectively, and $D$ is the mass diffusivity of the deficient reactant.  If the mass diffusivity of the limiting reactant is sufficiently larger than the thermal diffusivity, then the the Lewis number will be less than unity and below the critical value, and a cellular flame will be observed.

However, cellular structure has also been observed for flames where the heavier reactant is deficient, i.e., $Le>1$.  In this case, the flame is subject to the second type of instability that is hydrodynamic in nature.  A simplified view of a flame front is that it is a wave of density discontinuity that initially propagates at a constant speed, and the hydrodynamic instability is a result of the influence of the gas expansion on perturbations of the flame surface \citep{BechtoldMatalon1987,Bradley1999}.  Hydrodynamic instabilities arise for long wavelength perturbations, then shorter wavelength perturbations develop on the surface of the longer perturbations and also become unstable \citep{Bradley1999}.  The instabilities continue to cascade down to smaller and smaller wavelengths until the smallest wavelengths are stabilized by the thermodiffusive effects \citep{Bradley1999}.

Flame stability analysis of spherical flames \citep{BechtoldMatalon1987,BradleyHarper1994} results in an expression for the growth rate of an instability in terms of series of spherical harmonic integers, $n$, which are related to the wavenumber.  The growth rate of the amplitude of a flame surface perturbation with spherical harmonic integer $n$, $\bar{A}\left(n\right)$, is found to be
\begin{align}
\bar{A}\left(n\right) &= \omega - \omega\frac{\Omega}{Pe} \; .
\end{align}
The first term, $\omega$, represents the contribution to the growth rate due to the hydrodynamic instability, and the second term, $\omega \Omega/Pe$ is the contribution due to thermodiffusive effects.  The hydrodynamic term $\omega$ is a function only of $n$ and the flame expansion ratio $\sigma = \rho_u / \rho_b$ where $\rho_u$ and $\rho_b$ are the densities of the unburned and burned gases, respectively.  In the thermodiffusive term, $\Omega$ is a function of $\omega$ (and hence $n$ and $\sigma$) and the Markstein number, $Ma$, which describes the flame response to stretch.  The Peclet number, $Pe$, is defined as the ratio of the flame radius to the flame thickness,
\begin{align}
Pe &= \frac{r}{\delta_l} \; .
\end{align}
When the flame radius is small, the Peclet number is small and the flame stretch rate is high enough to maintain stability \citep{Bradley1999}.  As the flame propagates, a critical Peclet number, $Pe_c$, is reached (with a critical flame radius $r_c$) and the growth rate of an instability at a critical wavelength $\lambda_c$ becomes positive.  Therefore, at this critical radius the onset of instability is observed and the flame attains a cellular structure and smaller and smaller cells are generated.  However, the flame stretch rate restabilizes the instabilities, causing the cells to increase in area until a second critical Peclet number, $Pe_{cl}$, is attained \citep{Bradley1999}.  At this point the flame stretch is no longer able to stabilize the flame and cells are generated at smaller and smaller wavelengths, with the shortest wavelength in the cascade associated with stabilization of the hydrodynamic instability by thermodiffusive effects.  Therefore, the instability of a spherically expanding flame front is controlled by competition between hydrodynamic instability, flame stretch, and diffusion.

\section{Unstable Flame in a Lean H\texorpdfstring{$_2$}{H2}-N\texorpdfstring{$_2$}{N2}O Mixture}

This fluid dynamics video shows ignition and propagation of an unstable flame in a gaseous mixture of hydrogen (H$_2$) and nitrous oxide (N$_2$O).  The composition is fuel-lean, with equivalence ratio $\phi=0.40$, where
\begin{align}
\phi = \frac{x_{fuel}/x_{oxidizer}}{\left(x_{fuel}/x_{oxidizer}\right)_{stoic}} &= \frac{x_{H_2}/x_{N_2O}}{1}
\end{align}
and $x$ is the number of moles.  The initial temperature and pressure of the gas mixture was 297 K and 40 kPa, respectively.  The Lewis number of the mixture was less than 0.30 and below the critical Lewis number, and so the flame was unstable to thermodiffusive effects at small flame radii.  The visualization of the flame propagation was achieved using schlieren optics and a high-speed camera.  The flame was ignited using a capacitive spark discharge between two electrodes made of tungsten wire.

The video begins by showing the hot gas produced by the spark and the shock wave that is emitted by the expanding gas kernel.  An approximately spherical flame forms, and perturbations are initiated in the flame front by the electrodes.  The perturbations appear as ``cracks'' on the flame surface, and due to thermodiffusive instability the cracks propagate across the surface with a speed related to the growth rate $\bar{A}\left(n\right)$.  As multiple cracks propagate across the flame surface, cross-cracking between the instabilities also occurs.  Eventually, the critical Peclet number is reached and hydrodynamic instability contributes to the perturbation growth.  

By 3 ms after ignition, the flame surface is completely covered by the cellular structure.  Smaller and smaller cells are formed as shorter wavelength instabilities begin to grow until the cells are stabilized by the flame stretch rate, causing an increase in the cell area.  As the flame continues to propagate, the stabilizing affect of stretch is diminished and the second critical Peclet number is reached.  Smaller cells form once again on the surface of the larger instabilities and are subsequently stabilized by thermodiffusive effects.  However, in the video the cell areas are shown to increase again before breaking down into even smaller cells.  The competition between the hydrodynamic instability and the stabilizing effects of flame stretch and diffusion lead to alternating growth and splitting of cells on the flame surface several times over, lending a pulsating appearance to the flame.

\section{Conclusion}

Ignition and flame propagation in hydrogen in hydrogen-nitrous oxide mixtures is a topic of significant importance for safety in nuclear waste storage and semi-conductor manufacturing \citep{MevelEtAl2009}.  Understanding the unstable nature of the flames is important because instabilities cause acceleration of the flame front, which can lead to deflagration-to-detonation transition.  If the flame transitions to a detonation, extremely high pressures are generated that can pose risks both to structures and human safety.  This fluid dynamics video shows that H$_2$-N$_2$O flames can be extremely unstable and exhibit the interesting feature of having the cells on the flame surface grow in area then split into ever smaller cells multiple times.  The video also shows that these instabilities cause the flame to accelerate significantly.  Future work will involve using flame stability theory to describe and predict the features of instabilities in H$_2$-N$_2$O flames.

\end{document}